\documentclass[prl,twocolumn,aps]{revtex4}
\usepackage{graphicx}
\usepackage{amsfonts}

\newcommand{\be}{\begin{equation}}
\newcommand{\ee}{\end{equation}}
\newcommand{\bea}{\begin{eqnarray}}
\newcommand{\eea}{\end{eqnarray}}
\newcommand{\beq}{\begin{equation}}
\newcommand{\eeq}{\end{equation}}
\newcommand{\beqn}{\begin{eqnarray}}

\newcommand{\eeqn}{\end{eqnarray}}

\usepackage{amsfonts}
\usepackage{amssymb}
\usepackage{amsmath}
\usepackage{dsfont}
\usepackage{hyperref}
\usepackage{slashed}
\usepackage{empheq}
\usepackage{multirow}
\usepackage{fancyhdr}
\usepackage{appendix}
\usepackage{subfigure}

\begin{document}

\title{Rapid Thermalization in Field Theory  from Gravitational Collapse}
\author{David Garfinkle}\email{garfinkl@oakland.edu}
\affiliation{Department of Physics, Oakland University, Rochester, MI 48309}
\affiliation{Michigan Center for Theoretical Physics, Dept. of Physics, University of Michigan, Ann Arbor, MI 48109}
\author{Leopoldo A. Pando Zayas}\email{lpandoz@umich.edu}
\affiliation{Michigan Center for Theoretical Physics, Dept. of Physics, University of Michigan, Ann Arbor, MI 48109}

%\rightline{VPI-IPNAS-09-04}

%\renewcommand{\thefootnote}{\fnsymbol{footnote}}
%\date{\today}
\begin{abstract}
Motivated by the duality with thermalization in field theory, we study gravitational collapse of a minimally coupled massless scalar field in Einstein gravity with a negative cosmological constant. We investigate the system numerically and  establish that for small values of the initial amplitude of the  scalar field there is no black hole formation, rather,  the scalar field performs an oscillatory motion typical of geodesics in AdS. For large enough values of the amplitude of the scalar field we find black hole formation which we detect numerically as the emergence of an apparent horizon. Using the time of formation as an estimate for  thermalization in the field theory we conclude that thermalization occurs very rapidly, close to
the causal bound for a very wide range of black hole masses. We further study the thermalization time in more detail as a function of
the amplitude and the width of the initial Gaussian scalar field profile  and detect a rather mild structure.
\end{abstract}

\maketitle

%%%%%%%%%%%%%%%%%%%%%%%%%%%%%%%%%%%%%%%%%%%%%%%%%%%%%%%%%%%%%%%
{\it \large Introduction -- }
%%%%%%%%%%%%%%%%%%%%%%%%%%%%%%%%%%%%%%%%%%%%%%%%%%%%%%%%%%%%%%%
Understanding cooperative phenomena far from equilibrium poses one of the most challenging problems of present physics.  One  source of methods is  provided by the AdS/CFT correspondence  \cite{Maldacena:1997re} which  identifies a field theory without gravity with a string theory with gravity. In the classical limit of string theory one can use the appropriate generalization of Einstein's equations to follow the evolution in time of the fields in the dual field theory. In this context the gauge/gravity duality opens a particularly important window in  the absence of regular field theoretic methods: to study far from equilibrium phenomena one needs to study the evolution of Einstein's equation with appropriate boundary conditions.

In the framework of the AdS/CFT correspondence a field theory in equilibrium at finite temperature is dual to a black hole in asymptotically $AdS$ spacetime \cite{Witten:1998zw}. One very important development has been the establishment of the correspondence for applications to linear response theory for the near equilibrium region, that is, in the regime of  long wave lengths  and low energies with local fluid variables varying very slowly compared to microscopic scales (see \cite{Son:2007vk} for a review). The next frontier comes from the fact that the  evolution of spacetimes with the formation of black hole horizons is equivalent to non-equilibrium dynamics and evolution towards thermalization in field theory.

Besides the general reasons to study far from equilibrium phenomena using the gauge/gravity correspondence one practical motivation comes from the RHIC  and LHC experiments. Two crucial points are worth highlighting: (i) the observed Quark-Gluon Plasma (QGP) is strongly coupled and (ii) hydrodynamical description  fits in a wide range of processes.  Theoretical and experimental developments
indicate that the QGP produced at RHIC is a strongly interacting liquid rather than the weakly interacting gas of quarks and
gluons that was previously expected  \cite{Shuryak:2003xe,Shuryak:2004cy,Heinz:2004pj}. Another piece of evidence pointing to the need for methods applicable to strong coupled theories comes from the fact that the produced plasma locally isotropizes over a time scale of $\tau_{iso}\le 1 fm/c$. The dynamics of such rapid isotropization in a far-from-equilibrium non-Abelian plasma can not be described with the standard methods of field theory or hydrodynamics \cite{Muller:2008zzm}. In this letter we propose to study such rapid thermalization via its gravity dual -- gravitational collapse.

The question of thermalization is also crucial in condensed matter systems. There is an active interest in the understanding of the time evolution of a system following a quench \cite{RigolNature,PhysRevLett.103.100403,Calabrese:2006rx,Cardy:2011zz}. A holographic approach to this area is showing promise in lower dimensions \cite{AbajoArrastia:2010yt}. String theory also provides a framework for understanding some superconductors holographically \cite{Hartnoll:2008vx,Gubser:2008zu}. Moreover, some hard to understand properties of condensed matter system like the structure of non-Fermi liquids have been recently described using holographic models \cite{Cubrovic:2009ye} \cite{Faulkner:2010zz}. These models always involve a black hole and we believe that the formation of such a black hole will help our understanding of those systems at a more fundamental level.

In this paper  we present a full numerical analysis of the gravitational collapse of a massless minimally coupled scalar field in the presence of a negative cosmological constant. We go beyond previous attempts within the AdS/CFT approach that relied on perturbation theory \cite{Bhattacharyya:2009uu} or toy models for quench  \cite{Balasubramanian:2010ce,Erdmenger:2011jb,Balasubramanian:2011ur}. Our focus is on properties of the
thermalization time.

An intuitive explanation for the rapid thermalization time at RHIC within the context of the gauge/gravity correspondence is as follows.
Local gauge-invariant operators are mostly sensitive to geometry near the boundary. As explained in \cite{Bhattacharyya:2009uu}, in the large $N$ limit which is the region explored by the gauge/gravity correspondence, trace factorization ensures that the expectation value of products equals the product of expectation values. Only one-point functions of gauge invariant operators survive. Other studies supporting rapid thermalization in the context of the gauge/gravity correspondence include \cite{Balasubramanian:2010ce,Balasubramanian:2011ur,Ebrahim:2010ra,Asplund:2011qj}.

%%%%%%%%%%%%%%%%%%%%%%%%%%%%%%%%%%%%%%%%%%%%%%%%%%%%%%%%%%%%%%%%%%%%%%%%%
{\it \large Field Theory Setup -- }
%%%%%%%%%%%%%%%%%%%%%%%%%%%%%%%%%%%%%%%%%%%%%%%%%%%%%%%%%%%%%%%%%%%%%%%%
Our field theory setup is rather specific and we clarify it now. The overall field theoretic question one is interested in answering is: How does a field theory react to a rapid injection of energy? This is precisely what the RHIC experiment is all about for large values of the energy of the colliding particles -- How does QCD matter behave under such collision? Now the collision is clearly anisotropic as one has two gold atoms colliding. A gravity approximation to such experiment has been developed in the context of numerical relativity by Chesler and Yaffe in \cite{Chesler:2008hg,Chesler:2009cy,Chesler:2010bi}. A natural time scale in such collision process is the isotropization time which in the case of RHIC is $\tau_{iso}\le 1 fm/c$ \cite{Muller:2008zzm}. Our setup has one major difference. Namely, our injection of energy is spatially homogeneous, that is, it is the same in all points of the field theory space at a given time. We thus study a situation that is similar but not exactly equal to a quench. In this sense our setup is perhaps less directly applicable to RHIC-type experiments of collision but rather speaks loudly about more universal properties of strongly coupled field theories at large $N$ which are dual to gravity theories. What we set out to study in this paper, using gravity methods, is the time in field theory between the injection of energy at $t=0$  and the formation of the quark gluon plasma. We call this time the thermalization time. Of course, other interesting time scales might be present in a given experimental setup. One example is the above mentioned isotropization time, another natural time is the time at which the hydrodynamical approximation becomes a valid description of the quark gluon plasma. The thermalization time, as defined in our context, is expected to be of the same order but slightly smaller than these other two.

%%%%%%%%%%%%%%%%%%%%%%%%%%%%%%%%%%%%%%%%%%%%%%%%%%%%%%%%%%%%%%%%%%%%%%%%%
{\it \large Collapse in asymptotically $AdS_{5}$ spaces -- }
%%%%%%%%%%%%%%%%%%%%%%%%%%%%%%%%%%%%%%%%%%%%%%%%%%%%%%%%%%%%%%%%%%%%%%%%
We consider a minimal Einstein-scalar field action with a negative cosmological constant $\Lambda = -6/L^2$:

\be
\label{eq:Action}
S=\int d^{5}x \sqrt{-g}\bigg[\frac{1}{2\kappa}\left(R+\frac{12}{L^2}\right) -\frac12 (\partial\phi)^2 -U(\phi)\bigg].
\ee
We tackle this system using numerical methods developed and tested in \cite{Garfinkle:2004pw}, \cite{Garfinkle:2004sx} which we will describe in more detail below. We choose the metric Ansatz to be
\be
ds^2 = - \alpha^2 dt^2+ a^2 dr^2 + r^2 (d \chi^2 +\sin^2 \chi (d\theta^2 +\sin^2\theta d\varphi^2 )),
\ee
where $\alpha$ and $a$ are each functions of only $t$ and $r$. For convenience, let us define
\be
X=\partial_r \phi, \qquad Y=\frac{a}{\alpha}\partial_t \phi.
\ee
The Einstein equations can be written as:
\bea
\partial_r a =\frac{a}{r}(1-a^2)&+&\frac{r\, a}{6}\bigg[X^2+Y^2+a^2(U-\frac{12}{L^2})\bigg], \label{eq:a-prime} \\
\partial_r\ln (a\, \alpha)&=&\frac{r}{3}\left(X^2+Y^2\right), \label{eq:a-alpha-prime} \\
\frac{3}{r}\partial_t a&=&\alpha\, X\, Y. \label{eq:dot-a}
\eea
Here we are using units where $\kappa =1$.
It is worth pointing out that  equation (\ref{eq:dot-a}) can be shown to be  automatically satisfied when the other equations are satisfied.

The Klein-Gordon equation  takes the form
\be
\label{eq:Klein-Gordon}
\partial_t Y=\frac{1}{r^3}\partial_r\left(r^3\frac{\alpha}{a}X\right)-\alpha\, a\, \partial_{\phi}U.
\ee
 The main difference of the system given by equations (\ref{eq:a-prime}, \ref{eq:a-alpha-prime} and \ref{eq:Klein-Gordon}) with respect to the treatment presented in \cite{Garfinkle:2004pw} and \cite{Garfinkle:2004sx} lies in the powers of $r$ that appear. For example, in equation (\ref{eq:Klein-Gordon}) we have a cubic power or $r$ rather than a square one. This higher power of $r$ leads to a more singular behavior near $r=0$ which presents a numerical challenge.

%%%%%%%%%%%%%%%%%%%%%%%%%%%%%%%%%%%%%%%%%%%%%%%%%%%%
{\it \large Results -- }
%%%%%%%%%%%%%%%%%%%%%%%%%%%%%%%%%%%%%%%%%%%%%%%%%%%%%%%%%%%
We consider a massless scalar field, {\it i.e.}, $U=0$. We choose initial data of the form $\phi = A \exp (- {{(r- {r_0})}^2}/ \sigma^2 )$ where the amplitude $A$, center of the scalar
field profile $r_0$ and profile width $\sigma$ are constants.  This determines $X$ through $X={\partial _r}\phi$ and we choose
$Y=X$ initially so that the wave starts out purely ingoing.

\begin{figure}
\includegraphics[width=3.0in]{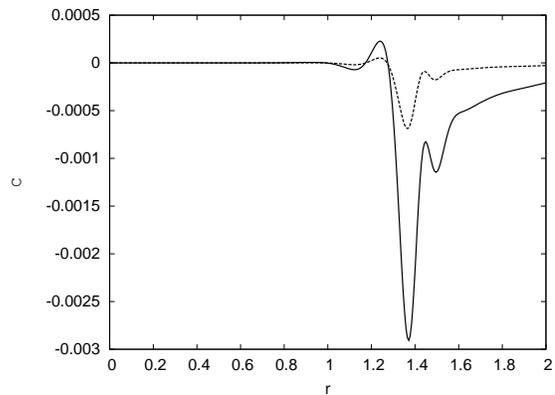}
\caption{\label{cnstrfig} The constraint quantity at coarse resolution (solid line) and fine resolution (dotted line). }
\end{figure}

We choose $L=1$ and for the spatial grid to be $0 \le r \le {r_{\rm max}}$ where $r_{\rm max}$ is a constant.  For the simulations
done in this work we use ${r_{\rm max}}=10$.  Since in anti-deSitter spacetime, more resolution is needed at small $r$ we do not
choose the spatial points to be evenly spaced.  Rather, for a simulation with $n$ spatial points, the value of $r$ at grid point
$i$ is ${r_i} = \sinh (k (i-1))$ where $k$ is a constant chosen so that ${r_n} = {r_{\rm max}}$.

We tested that the code is working properly by considering the constraint quantity
${\cal C} \equiv {\partial _t} a - r \alpha X Y /3$.  Note that it follows from eqn. (\ref{eq:dot-a}) that $\cal C$ vanishes.
However, the finite size of the grid spacing means that $\cal C$ will not vanish exactly in the simulation, but that instead in
a properly working simulation, a
smaller grid spacing should lead to a smaller $\cal C$.  In fig. (\ref{cnstrfig}) the result of a test of this sort is shown.
The solid line represents $\cal C$ for a simulation with $800$ grid points, while the dotted line is a simulation with $1600$ grid
points.  In both cases the simulations have $A=0.02, \, {r_0}=4.0$ and $\sigma = 1.5$ and the simulations are run to a time of
$t=0.454$.

\begin{figure}
\includegraphics[width=3.0in]{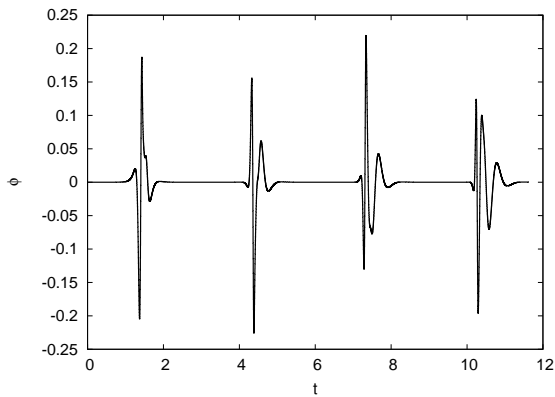}
\caption{\label{phi0fig} $\phi(0,t)$ for a small amplitude simulation. }
\end{figure}

For collapse in asymptotically flat spacetimes, small amplitude initial data leads to a wave that is initially ingoing, then
undergoes interference near the center and becomes an outgoing wave.  For collapse in asymptotically anti-deSitter spacetimes,
we would expect similar behavior, except that the outgoing wave should then reach anti-deSitter infinity, bounce and become
ingoing again, leading to another bounce, and so on.  And indeed, this is what we find.  Fig. (\ref{phi0fig}) shows the value
of the scalar field at the center as a function of time.  ($\phi (0,t)$).  The parameters for this simulation are
$A=0.0002, \, {r_0} =4.0$ and $\sigma =1.5$.  This simulation was done with $6400$ grid points.  Note that there are particular
periods of time where the scalar field at the center is non-negligible and that the middle of each such
time period is separated from the
next one by approximately $\pi$.  This is exactly what we would expect if the dynamics is mostly that of a massless scalar field
on a background anti-deSitter spacetime.  In the geometric optics limit, such a scalar field propagates along null geodesics, and
for anti-deSitter spacetime (with $L=1$) it takes a null geodesic a time of $\pi/2$ to propagate from the center to infinity.

For sufficiently large amplitudes the collapse process results in the formation of a black hole.  This is signalled by the
formation of a marginally outer trapped surface (also called an apparent horizon),
where the outgoing null geodesics cease to diverge from each other and instead
begin to converge.  In spherically symmetric spacetimes, an apparent horizon has the property that
$\nabla^a r \nabla_a r=0$.  In our coordinate system this would mean $a^{-2}\to 0$.  More precisely, our coordinate system
breaks down when an apparent horizon forms, and so our signal that a black hole is forming is that
$a \to \infty$ in the simulations.  Actually, it suffices to set a fairly moderate value for the maximum allowed value of $a$,
which we denote $a_{\rm max}$ and to stop the simulation whenever $a$ reaches $a_{\rm max}$, noting that a black hole has formed
at that time.  The position $r$ at which $a={a_{\rm max}}$ gives us the size of the black hole and allows us to estimate its mass.

We would like to know how long the process of black hole formation takes, {\it i. e.},  we would like to know how much time elapses
between the initial time and the time a marginally outer trapped surface forms.  For simulations it is natural to use the
coordinate $t$ as a measure of time, and this is the notion of time that we will use here.  In particular, we denote by
$t_{\rm AH}$ the coordinate time $t$ at which an apparent horizon first forms.
Note that our coordinate $t$ has geometric meaning in that it is the coordinate that is orthogonal to the
area coordinate $r$.  However, one might also want to know how long the process takes in terms of an ingoing null coordinate
$v$ as used in \cite{Bhattacharyya:2009uu}.  Such a coordinate is constant along ingoing null geodesics and can be normalized
by {\it e.g.} choosing it to be equal to the anti-deSitter time at infinity.

\begin{figure}
\includegraphics[width=3.0in]{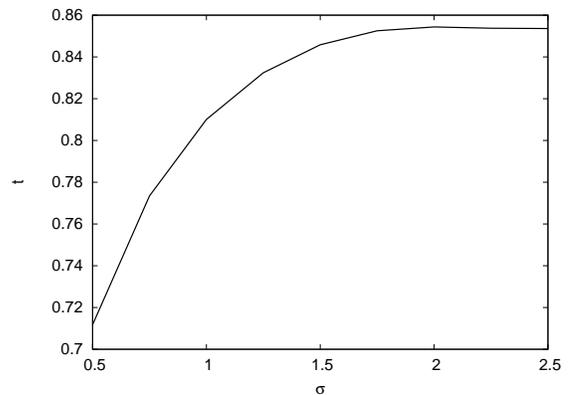}
\caption{\label{sigmatime} The dependence of $t_{\rm AH}$ on $\sigma$ }
\end{figure}

The time of black hole formation $t_{\rm AH}$ depends on the choice of initial data.  In particular, we want to know how
this time depends on the initial width of the pulse $\sigma$. This dependence is shown in fig. (\ref{sigmatime}) which gives
$t_{\rm AH}$ for several simulations with different values of $\sigma$.  For each of these simulations we have
$A=0.02$ and ${r_0} =4.0$.  These simulations were done with $6400$ grid points.

\begin{figure}
\includegraphics[width=3.0in]{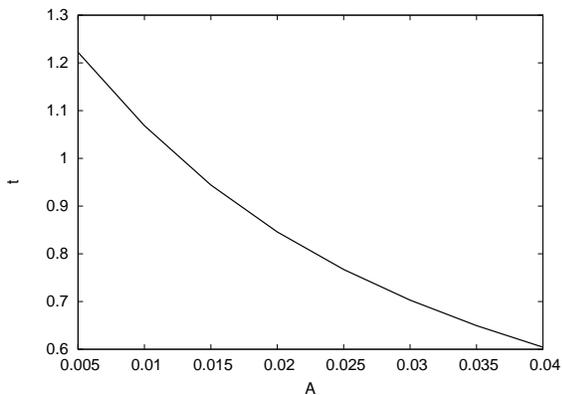}
\caption{\label{Atime} The dependence of $t_{\rm AH}$ on $A$ }
\end{figure}

We also want to know how
the time of apparent horizon formation depends on the initial amplitude of the pulse $A$.
This dependence is shown in fig. (\ref{Atime}) which gives
$t_{\rm AH}$ for several simulations with different values of $A$.  For each of these simulations we have
$\sigma =1.5$ and ${r_0} =4.0$. These simulations were done with $6400$ grid points.   The masses of the
black holes formed in the simulations range from $1.0$ for the smallest amplitude to $62.1$ for the largest amplitude, that is, over two orders of magnitude.

Some aspects of fig. (\ref{Atime}) can be understood using the properties of null geodesics in
anti-deSitter spacetime.  For a shell starting at infinity and propagating along a null geodesic, it would take a time
of $\pi/2$ to reach the center.  However, we start our shells at ${r_0}=4$ rather than infinity, and a null geodesic
takes a time of $t=1.32$ to propagate from $r=4$ to the center.  A shell of small mass must reach a small radius before
it forms an apparent horizon, so it is not surprising that at small amplitude $t_{\rm AH} \approx 1.32$.  However, the more massive a shell is, the larger the radius at which an apparent horizon forms, and therefore the shorter the time needed to propagate to that
radius.  Therefore we would expect $t_{\rm AH}$ to be a decreasing function of $A$, and fig. (\ref{Atime}) confirms this expectation.

%%%%%%%%%%%%%%%%%%%%%%%%%%%%%%%%%%%%%%%%%%%%%%%%%%%%%%%%%%%%%%%%%%%%%%%%%%%%%%%%%%%%%%%%%%%%%%%%%%%%%%%%%%%%%%%%%%%%%%%%%%
{\it \large The thin Vaidya shell approximation -- }
Of the many setups discussed in the literature, the one that can be most directly compared to our work here is the one presented in \cite{Bhattacharyya:2009uu}. This work provided, to our knowledge, the first systematic analysis of the thin shell Vaidya collapse as the dual of rapid thermalization in field theory.

Let us briefly discussed the setup in \cite{Bhattacharyya:2009uu} which  considered an action of the form (\ref{eq:Action}).
Their metric is spherically symmetric and is given in Vaidya type coordinates
\bea
ds^2&=&2dr dv -g(r,v)dv^2 +f^2(r,v)d {\Omega ^2}
\nonumber \\
\phi &=& \phi(r,v).
\eea
where $d {\Omega ^2}$ is the line element of the unit $S^3$. The crucial physical information is stored in the scalar field profile which is $\phi_0(v) <\epsilon$ for $0<v<T$ and otherwise vanishes.
In \cite{Bhattacharyya:2009uu}, a perturbation theory in $\epsilon$ was developed.  The situation we treat is not precisely
the same as that of \cite{Bhattacharyya:2009uu}: our initial data is different, and the fact that we use different coordinate
systems makes direct comparisons somewhat involved.  Nonetheless, there are many similarities.  Our Gaussian profile falls off
so fast that it might as well be of compact support, and our choice of ingoing waves means that at large $r$ the profile is
essentially a function of an ingoing null coordinate $v$.  We can also choose our parameters $A$ and $\sigma$ (corresponding
respectively to the $\epsilon$ and $T$ of \cite{Bhattacharyya:2009uu}) sufficiently small to be within the perturbative
regime of \cite{Bhattacharyya:2009uu} or sufficiently large that that perturbative regime is no longer valid.
Furthermore, in both cases many of the main features of the collapse
process seem to depend simply on the approximate propagation of the scalar field along null geodesics until the shell becomes
sufficiently small that a trapped surface can form.

The main advantage of our method is that we can state rather precisely such things as
when a black hole forms and how large it is without having
to be in the regime where the perturbation expansion of \cite{Bhattacharyya:2009uu} is well approximated by its first
couple of terms.

%

%%%%%%%%%%%%%%%%%%%%%%%%%%%%%%%%%%%%%%%%%%%%%%%%%%%%%%%
{\it \large Conclusions -- }
%%%%%%%%%%%%%%%%%%%%%%%%%%%%%%%%%%%%%%%%%%%%%%%%%%%%%%%
Stated in terms of the coordinate time used in our simulations, the time of
black hole formation is not instantaneous and depends on the time the shell takes to
propagate to a sufficiently small radius that its mass will give rise to a horizon.  However,
because a sufficiently thin shell propagates along an ingoing null geodesic, this means that
if an ingoing null coordinate is chosen as the time, then black hole formation is essentially
instantaneous.  Thus the question of rapid thermalization in the AdS/CFT correspondence seems to
hinge on the question of what bulk coordinate is the dictionary translation of time in the boundary CFT. Our explicit simulations show
rapid thermalization with times always comparable to the time it takes a null geodesic to travel from the center of the shell profile
to the radius at which the shell is sufficiently compact to form a horizon. We obtain our results for masses ranging over two orders of magnitude. In a more detailed study we verify that the dependence on the width is milder than the dependence on the amplitude.

We have laid the foundation that will allow us to provide, in full detail, various other properties of the thermalization process by studying its
gravitational dual. For example, the role of a  mass term in the potential for the
scalar field. In the AdS/CFT correspondence the mass of the scalar field is related to the conformal dimension of the dual
operator. We have considered a massless
field corresponding to a dimension four operator in the field theory. Following the discussions presented in
\cite{Bhattacharyya:2009uu,Balasubramanian:2010ce,Balasubramanian:2011ur,Ebrahim:2010ra}, we plan to discuss two-point functions, Wilson loops and the
entanglement entropy in our collapsing simulations. The study of these quantities naturally allows for a more precise definition of thermalization time than the general one used in this paper. We will present those results elsewhere.

%%%%%%%%%%%%%%%%%%%%%%%%%%%%%%%%%%%%%%%%%%%%%%%%%%%%%%%
{\it \large Acknowledgments -- }
%%%%%%%%%%%%%%%%%%%%%%%%%%%%%%%%%%%%%%%%%%%%%%%%%%%%%%%
L. PZ is grateful to H. de Oliveira and C. Terrero-Escalante for collaboration on similar matters. We thank D. Reichmann, D. Trancanelli for useful conversations. This work is  partially supported by Department of Energy under grant DE-FG02-95ER40899 to the University of Michigan and by NSF grant PHY-0855532 to Oakland University.

% If you don't have the corresponding .bst and .bib files, comment the two lines below and copy paste the content of the .bbl file
%there (whoever compiled the %bibliography should send you the .bbl file)
%\bibliographystyle{JHEP}
%\bibliography{Collapsebib}

\end{document}